\documentclass[pre,twocolumn,aps,showpacs,amsmath,amssymb,floatfix]{revtex4}

\usepackage{graphicx}

\begin{document}


\title{A study of a local Monte Carlo technique for simulating systems of charged particles}

\author{P. A. McClarty}
\affiliation{School of Physics and Astronomy, University of Manchester, Manchester M13 9PL, U.K.}  

\date{\today}

\begin{abstract}
We study some aspects of a Monte Carlo method invented by Maggs and Rossetto for simulating systems of charged particles. It has
the feature that the discretized electric field is updated locally when charges move. Results of simulations of the two dimensional one-component plasma are presented. Highly accurate results can be obtained very efficiently using this lattice method over a large temperature range. The method differs from global methods in having additional degrees of freedom which leads to the question of how a faster method can result. We argue that efficient sampling depends on charge mobility and find that the mobility is close to maximum for a low rate of independent plaquette updates for intermediate temperatures. We present a simple model to account for this behavior. We also report on the role of uniform electric field sampling using this method.    
\end{abstract}


\maketitle

\section{Introduction} \label{introduction}

As an aid to understanding the behavior of statistical systems, it is
common practice to carry out computer simulations of models that are invented to capture the
physical behavior of interest because, even for
unelaborate models, it is often extremely difficult to find approximate
and especially exact analytical results. Both Monte Carlo and
molecular dynamics methods work by generating large numbers of
particle or field configurations from which correlation functions can
be obtained that converge to thermal averages by design. In Monte
Carlo, each new configuration is produced by making a pseudorandom
change to the last one in the chain, testing the ratio of Boltzmann
weights (which, in the canonical ensemble, amounts to a computation of
the energy difference), and choosing either the new configuration or
the old one. Molecular dynamics is a numerical evolution of a many-body
system from the equations of motion of the component 'particles' with
some constraint depending on the ensemble - for example to maintain a
constant temperature. The degree of approximation to thermal
equilibrium in both cases is limited, in practice, by the efficiency
with which the algorithm samples the regions of highest probability in
the space of configurations, and the rate at which configurations can
be produced. These two factors determine the range of system sizes
that can be sampled effectively, though a compromise must often be
made between obtaining large numbers of configurations based on the
relaxation time of the combined model and algorithm and choosing large enough systems to lessen finite size effects that obscure the
thermal equilibrium results required and which are usually poorly understood.

This work is a study of a Monte Carlo technique for simulating systems of thermalized classical charges. Most of the systems
studied in soft condensed matter laboratories fall into this class. It includes simple liquids, polymers, colloids, liquid
crystals, proteins and other biologically active molecules. 

In many interesting cases, charged particle interactions can be
screened by the charges themselves, giving an effective interaction of the Yukawa
form with a screening length that can often be tuned in
experiments. But there are also many cases where one must consider the
exact Coulomb interaction. Because the Coulomb interaction is long-ranged,
one must presumably either compute all pairwise interactions with the charge that
is moved at a given step (in Monte Carlo) or one must solve the
Poisson equation at each step. Consider a Monte Carlo simulation with $N$
charges. When a charge is moved, the energy difference between
configurations requires $O(N)$ operations for all pairwise
interactions. Each Monte Carlo \emph{sweep}, which is defined to be $N$ charge moves or Monte Carlo \emph{steps}, requires $O(N^{2})$ operations;
this is how the simulation will slow down as the system size increases. With short range interactions, in contrast, we expect
$O(N)$ operations for each sweep.

The order $O(N^{2})$ scaling for Coulomb interactions is a real
problem and a great deal of effort has been devoted to improving the
scaling and the prefactor that accompanies it. This study focuses on a relatively
new method proposed by Maggs and Rossetto \cite{RossettoMaggs}, which achieves $O(N)$
scaling and which has some promise as a competitive approach for
simulating charged particles using either molecular dynamics or Monte Carlo.

\subsection{Local Electrostatics} \label{averaging}

The gist of the method of Maggs and Rossetto is as follows. Rather
than solving the Poisson equation (which follows from Gauss' law
supplemented with the condition that the electric field circulation vanish) at each step, one imposes only
Gauss' law. The electric field circulation decouples from the charges, so it
averages out during the simulation and one obtains electrostatic averages. Gauss' law can be imposed
locally in the sense that when a charge moves, this divergence condition can be maintained by changing the electric field in the
neighborhood of the charge only. Transposing this result to Monte Carlo, we see that each sweep requires only $O(N)$ operations.

Now we fill in details of the argument \cite{RossettoMaggs}. Consider a number
of charged particles with charge density $\rho(\mathbf{r})=
\sum_{i}q_{i}\delta(\mathbf{r}-\mathbf{r}_{i})$. Suppose the electric field $\mathbf{E}$ is affected by the charge distribution only
to the extent that Gauss' law is satisfied
\[ \mathbf{\nabla}\cdot\mathbf{E} = \rho(\mathbf{r}). \]
In other words we do not impose the further condition required for
electrostatics that the curl of the electric field be equal to zero. The
charged particle system is closed so that the amount of charge does
not change and the whole system is immersed in a heat bath at inverse temperature $\beta$. The partition function
in this canonical ensemble in two dimensions is
\begin{multline} Z(\beta) = \int \prod_{i}d^{2}\mathbf{r}_{i}\int
\mathcal{D}\mathbf{E} \, \delta(\mathbf{\nabla}\cdot\mathbf{E} -
\rho(\mathbf{r})) \\ \times \exp\left(-\frac{\beta}{2}\int d^{2}\mathbf{r}\,
\mathbf{E}^{2}(\mathbf{r}) \right), \label{eqn:partition}\end{multline}
in which the factor resulting from the momentum integration over each
particle has been left out. The electric field can be split into an
irrotational or longitudinal part $\mathbf{E}_{\|}$ and a transverse
part $\mathbf{E}_{\bot}$: $\mathbf{E} = \mathbf{E}_{\|} +
\mathbf{E}_{\bot}$ satisfying
\[ \begin{array}{ccc}\mathbf{\nabla}\cdot\mathbf{E}_{\bot} = 0 & \mbox{and}
& \mathbf{\nabla}\times\mathbf{E}_{\|} = 0. \end{array} \]
Gauss's law can now be written as $\mathbf{\nabla}\cdot\mathbf{E}_{\|}
= \rho(\mathbf{r})$ and with the condition on the longitudinal field
that its curl vanish, $\mathbf{E}_{\|}$ is derivable from a scalar
field $\phi(\mathbf{r})$ which obeys Poisson's equation
\begin{align*}  &\nabla^{2}\phi(\mathbf{r}) =  -\rho(\mathbf{r}) \\ & \mathbf{E}_{\|} =  -\mathbf{\nabla}\phi(\mathbf{r}). \end{align*}
The solution to the Poisson equation in two dimensions is uniquely 
\[ \phi(\mathbf{r}) = -\frac{1}{2\pi}\int d^{2}\mathbf{r}'
\rho(\mathbf{r}')\ln|\mathbf{r} - \mathbf{r}'|. \]
We return to the energy functional to see the effect of separating out
the longitudinal and transverse contributions
\[  \int d^{2}\mathbf{r} \, \mathbf{E}^{2}(\mathbf{r}) = \int
d^{2}\mathbf{r} \left[(\mathbf{\nabla}\phi)^{2} + \mathbf{E}_{\bot}^{2} -
  2\mathbf{\nabla}\phi\cdot\mathbf{E}_{\bot} \right]. \]
Partially integrating the third term, we get
\[ 2\int  d^{2}\mathbf{r} \,\phi(\mathbf{\nabla}\cdot\mathbf{E}_{\bot}) = 0. \]
The whole of the dependence of the energy on the charge positions is contained in
the scalar potential so the partition function factorizes:
\[ Z(\beta) = Z_{\mathbf{E}_{\bot}}(\beta) \times \int
\prod_{i}d^{2}\mathbf{r}_{i} \, \exp\left( -\frac{\beta}{2}\int d^{2}\mathbf{r} \,
(\mathbf{\nabla}\phi)^{2} \right) \]
where $Z_{\mathbf{E}_{\bot}}(\beta)$, which comes from the Gaussian integration over the transverse degrees of freedom, is independent of
the charge configuration. So thermodynamic averages derived from the
partition function (\ref{eqn:partition}) are the same as those that would be found from the
partition function with the electrostatic energy functional. 

Because Gauss' law can be imposed locally, we ought to be able to run Monte Carlo
simulations of systems of charged particles with local field updates
and hence local energy computations at each step rather than solving
Poisson's equation to find the new unique global field configuration
for each charge configuration. The advantage of having local energy computations at each step is an
improvement in the scaling of the algorithm over methods involving
direct evaluation of the pairwise interaction. Each particle move is a
local operation involving $O(1)$ operations, so the scaling of the time
per Monte Carlo sweep is $O(N)$. The transverse field must also be
sampled, and yet one finds that this does not contribute significantly to
the scaling. This latter fact is quite counterintuitive and we shall
devote some more attention to it in the following. 

It is worth pointing out that if we begin with Maxwell's
electrodynamics, a similar argument to the one given above shows that electrostatic
averages follow in this case. This is not remarkable given the
decoupling of the divergence and circulation of the electric field in
the pure electric field case described above, but we set out this result in an appendix. One could therefore devise a molecular dynamics algorithm
with the appropriate constraint for working at constant temperature
that integrates Maxwell's equations (which are local equations) at each time-step and with the speed of light as a variable,
chosen to be small enough so that the algorithm is truly local, and obtain the same results that would be obtained directly from
electrostatics. However, Maggs and Rossetto have shown that one is free to find the
simplest local dynamics possible for the charges that simultaneously samples longitudinal and transverse degrees of freedom
subject to Gauss' law, and use this dynamics to perform much more efficient Monte Carlo or molecular dynamics simulations than
would be possible with full Maxwell electrodynamics. In this paper, we shall use a charge and field dynamics invented by Rottler and
Maggs \cite{RottlerMaggs} that omits magnetic fields altogether.

\subsection{Overview} 

In the next section, we provide a very short introduction to charge interactions in periodic media in the continuum limit and for
the case where the electric field is discretized onto a lattice. Hence, we describe the conventional $O(N^{2})$ global Monte Carlo
method we have used and give a practical account of the corrections that must be made to lattice simulations to better approximate
the continuum pair-wise potential. Section~\ref{localmethod} describes the implementation of the local $O(N)$ method for Monte Carlo simulations
with charges moving off-lattice with particular attention to unusual features of the dynamics; I also present a continuum version
of the local update method which shows some already known features very transparently. Global and local Monte Carlo
methods are applied in Section~\ref{2DOCP} to the 2D OCP and we discuss the accuracy of the local methods and the effect of
uniform field sampling. Then, we shall see that charge autocorrelation times are much greater for the local method than for global
methods. Section~\ref{speed} includes the autocorrelation data into a comparison of the speeds of the local method and a global
method for typical system sizes. Results of an investigation into the effects of the transverse electric field on charge
relaxation times are given in Section~\ref{transautoc}. We also discuss the problem of how additional degrees of freedom result in
$O(N)$ scaling when the naive expectation is that transverse fields should be allowed to relax across the system at each step. The
first appendix derives a useful result for corrections to 2D lattice-based simulations. The second appendix shows that the
argument of Maggs and Rossetto given above carries over to Maxwell electrodynamics.  


\section{Simulation methods} \label{methods}

The Monte Carlo simulations that will be discussed below are based on the acceptance procedure of Metropolis and co-workers
\cite{Metropolis}. This involves the computation of a difference in energies of two field configurations at each step. The
simulation methods we have used are distinguished by the ways in which this difference is found. 

In this section, we discuss the interactions between charged particles in rectangular cells with periodic boundary conditions and
the corrections that have to be made to the interaction when the electric field is put onto a lattice in order to better approximate the continuum potential energy. 
We describe the simulation methods we have used in turn, starting with the continuum case, then the problems of discretization, and leaving the
special properties of the local algorithm until the next section.  In the following, unless otherwise stated, we consider two
dimensional electrostatics for notational simplicity and because this is most appropriate to the simulations that we have performed.

\subsection{Periodic systems}

The potential energy of a finite number of charges in a rectangular cell with periodic boundary conditions is a \emph{conditionally}
convergent series over all pairwise Coulomb energies for the charges in the infinite array of cells implied by the boundary
conditions. Each cell must be charge neutral otherwise the series is divergent. And if the dipole moment in each cell is zero the
sum is absolutely convergent. In general, the dipole moment in each cell is non-vanishing, so the problem is to devise a physically
motivated summation of the series. A further problem is to identify the unique solution to the Poisson equation with periodic
boundary conditions. These have been solved by Leeuw and Perram \cite{Leeuw2} by obtaining the potential for a large spherical cluster of
identical cells embedded in a medium of dielectric constant $\epsilon$; their expression includes a term that depends on the net dipole
moment of the elementary cell and the dielectric constant of the surrounding medium. The unique periodic solution to the Poisson
equation is obtained in the limit as $\epsilon\rightarrow\infty$ and corresponds to the case of a large spherical cluster of
identical rectangular cells of charges surrounded by a perfect conductor that perfectly cancels the dipole moment of the cluster
of cells \cite{Fraser}. The periodic potential, called the Ewald potential, ((\ref{eqn:ewald}) given in the first appendix for the
two dimensional case \cite{Leeuw2}) is a pairwise potential for charges in a rectangular box that includes the direct Coulomb
potential and also the potential due to all the image charges for the pair considered. The dipole moment term that is present for
finite $\epsilon$ takes the form
\begin{equation} U_{\mbox{dip}} = \frac{\pi}{2L_{x}L_{y}(\epsilon + 1)}\left( \sum_{a=1}^{N} q_{a}\mathbf{r}_{a} \right)^{2}
  \label{LPdipole} \end{equation}
for cell lengths $L_{x}$ and $L_{y}$ and with charge $q_{a}$ at position $\mathbf{r}_{a}$. This will be discussed further in connection with the local method.

The Ewald potential splits the calculation of the potential into short-range and long-range parts. The latter is performed in
Fourier space. This forms the basis for molecular dynamics simulations. However, the Ewald potential is not always the most
efficient starting point for computations. I have used the form found by Lekner \cite{Lekner1,Lekner2} and developed for
rectangular periodic cells by Gr\o nbech-Jensen \cite{GronJen}. If the charges are separated by a vector with Cartesian components
$x$ and $y$ and the cell lengths are, once again, $L_{x}$ and $L_{y}$, then the pairwise potential energy between unit charges is
\begin{multline}
 2\pi V  =  K\left(\frac{L_{y}}{L_{x}}\right) +
  \frac{\pi L_{y}}{L_{x}}\left(\frac{y}{ L_{y}}\right)^{2} \\ - \frac{1}{2}\ln\left\{ \prod_{k=-\infty}^{\infty}
  \frac{\cosh\left(2\pi\frac{L_{y}}{L_{x}}\left(\frac{y}{L_{y}}+k\right)\right)-\cos\left(\frac{2\pi x}{L_{x}}\right) }{\cosh\left(2\pi\frac{L_{y}}{L_{x}}k\right)}\right\}
\label{Lekner} \end{multline}
where $K$ is a constant that includes the interaction of the charges with the uniform background charge and with all their image
charges and is given by
\[ K\left(\frac{L_{y}}{L_{x}}\right) = \frac{\pi}{6}\frac{L_{y}}{L_{x}} - \frac{1}{2}\ln 2 - \ln \prod_{k=1}^{\infty}\left( 1 +
\exp\left(-4\pi \frac{L_{y}}{L_{x}}k\right)\right). \] The advantage of this formula is that it is in terms of elementary
functions; although it is an infinite product in its main part, the product converges very quickly - I take $|k|\leq 5$ which
agrees with the exact result to machine precision.

\subsection{Lattice simulations}

The $O(N)$ Monte Carlo method that is the subject of this paper requires that we discretize the electric field so that it lives on
the links of a square mesh with charges on the sites. Overall charge neutrality must be maintained (if necessary by having a uniform
background charge).  We consider the case of a two-dimensional square
lattice with lattice spacing $a$, site labels $\{ \mathbf{R} \equiv (n_{x},n_{y}) \}$ and link directions
 $\mu = x, y$ such that charge $q(\mathbf{R})$ is assigned (in the way described above) to site $\mathbf{R}$ and electric fields are defined on the
 links of the lattice with $E(\mathbf{R},-\mu) = -E(\mathbf{R}-\mu,\mu)$ by convention.  The lattice has $N_{x}=L_{x}/a$, $N_{y}=L_{y}/a$ plaquettes in the
$x$ and $y$ directions respectively. Sites $n_{x}=0$ and $n_{x}=N_{x}$ are identified. Periodicity in the $y$ direction is
implemented in the same way. 

The discrete form of Gauss' law is \begin{equation} \sum_{\mu=x,y} a ( E(\mathbf{R},\mu) +
  E(\mathbf{R},-\mu))  =  q(\mathbf{R}). \label{latticeGauss} \end{equation} 

When the Poisson equation is discretized onto the lattice, its solution can be written as
\[ \phi(\mathbf{R}) =  \sum_{\mathbf{R}'}G(\mathbf{R} - \mathbf{R}')q(\mathbf{R}'). \]  The
charge at lattice site  $\mathbf{R}$ is $q(\mathbf{R})$ and $G(\mathbf{R})=G(n_{x},n_{y})$ is the lattice Green's function. One can show, by
calculating the discrete Fourier transform of the Poisson equation with periodic boundary conditions, that
\begin{multline} G(n_{x},n_{y}) =\sum_{p_{x}=0}^{N_{x}-1} \exp\left(2\pi i
\frac{p_{x}n_{x}}{N_{x}}\right) \sum_{p_{y}=0}^{N_{y}-1} \exp\left(2\pi i
\frac{p_{y}n_{y}}{N_{y}}\right) \\ \times\frac{1}{N_{x}N_{y}}
\frac{1}{4-2\cos(\frac{2\pi p_{x}}{N_{x}}) - 2\cos(\frac{2\pi p_{y}}{N_{y}})} \label{Greensfn}
\end{multline} 
in two dimensions. The electric field is related to the potential on the lattice by $a
E_{x}(n_{x},n_{y})=\phi(n_{x}+1,n_{y})-\phi(n_{x},n_{y})$ and similarly for the $y$ component.

The charged particles are not confined to the lattice sites in our simulations. In common with most molecular dynamics simulations
of charged particles, charges are smoothly assigned to lattice sites in the neighborhood of each particle as a function of the
particle position. Suppose that each charge $q_{a}$ at continuous position $\mathbf{r}_{a}$ is
spread over lattice sites $\{\mathbf{R}_{i}\}$. The charge on site
$\mathbf{R}_{k}$ is given by $q_{a} s(\mathbf{r}_{a},\mathbf{R}_{k})$. We
call $s$ the charge assignment function which has the property 
\[ \sum_{j}s(\mathbf{r}_{a},\mathbf{R}_{j}) = 1. \] 
Many choices of assignment function are possible and the choice may greatly affect the performance of the algorithm. Clearly,
as the number of lattice sites over which each charge is distributed increases, the computational cost increases also. As for the
accuracy, the overlap of lattice based charge clouds results in a deviation from the continuum potential for charge separations on the scale of the
charge cloud. Because, for high accuracy, one should correct the energy for charges with overlapping charge clouds, there is a
greater computational cost for higher order assignment functions. For Fourier transform methods, including most molecular dynamics
methods \cite{Mesh}, accuracy can increase as the radius of the charge cloud increases. This is because narrow charge clouds have a large spread
in Fourier space and higher frequency modes can act as aliases for lower frequency ones in the sense that modes of different frequency can appear the same on the scale of the
lattice if their values at the lattice sites coincide. To avoid this aliasing error, it is better, for these simulation methods, to have larger charge clouds so
that frequency aliases have a smaller amplitude. We use a third order scheme due to Hockney and Eastwood \cite{HandE} that spreads the charges
over nine vertices in two dimensions. If $Dx$ and $Dy$ are the
distances in units of the lattice spacing $a$ of the charge $q$ to the
vertex $v_{0,0}$ closest to it, the weights for charge assignment in the $x$ direction are 
\begin{equation} \begin{array}{c} W^{x}_{-1} = (Dx - 0.5)^{2}/2 \\ W^{x}_{0} = 0.75 - Dx^{2} \\ W^{x}_{+1} = (Dx + 0.5)^{2}/2\end{array} \label{eqn:weights} \end{equation} 
with similar formulas for the $y$ direction assignment. The charge assignment function $s$ is a product of pairs of weights as shown in Figure~\ref{fig:path}.

The total energy of the system in terms of the Green's function is
\begin{equation} U =
\frac{1}{2}\sum_{\mathbf{r}_{a}}\sum_{\mathbf{r}_{b}}\sum_{i,j} q^{2} G(\mathbf{R}_{i} - \mathbf{R}_{j})s(\mathbf{r}_{a},\mathbf{R}_{i})s(\mathbf{r}_{b},\mathbf{R}_{j}).
\label{eqn:energy}
\end{equation}
Alternatively, it is the sum of all the squared link electric fields
\begin{equation} U = \frac{a^{2}}{2}\sum_{n_{x}=0}^{N_{x}-1}\sum_{n_{y}=0}^{N_{y}-1} \left( E^{2}((n_{x},n_{y}),x) +
  E^{2}((n_{x},n_{y}),y) \right). \label{eqn:energy2} \end{equation}

This interaction is the one generated by the local simulation
algorithm. At charge separations greater than about one lattice
spacing (depending on the charge assignment scheme used) this
interaction energy is an excellent approximation to the Ewald
potential. At short range the interaction is weaker than in the
continuum problem. In fact the limit of the interaction energy as the
separation of two charges tends to zero is finite. For accurate
simulation this must be corrected. For each charge move, charges in
the neighborhood of the charge that is being moved are identified using the linked list method of
Hockney and Eastwood \cite{HandE}. The discretized interaction energy from equations
(\ref{eqn:energy}) (with the sum suitably restricted) and (\ref{Greensfn}) is subtracted off for each
pair of neighboring charges and the Lekner potential $V$ from
equation (\ref{Lekner}) or a good short range approximation to the Ewald potential is
added on in its place. A suitable approximation for small charge
separations $r$ is 
\begin{equation} U_{SRC} = -\dfrac{q^{2}}{2\pi}\ln r + \dfrac{q^{2}r^{2}}{4A} \label{SRCmain} \end{equation}
where $A=L_{x}L_{y}$, which is derived in an appendix and presented with a calculation of the deviation from
the Lekner potential for different charge separations. For very high accuracy or for low temperatures,
the Lekner interaction is more appropriate. The linked list method is a local method so that the scaling of the local
algorithm is preserved. The periodic cell is divided up into chaining
cells - regions of $2\times2$ plaquettes in our simulations. At the
beginning of the simulation, the charges are given chaining cell labels
and this information is put into a linked list. The linked list is
composed of two arrays; the first gives a charge label given a
chaining cell label - in other words it gives the label of one charge
in a given chaining cell; the second array has particle labels as its
contents and also as its indices in such a way that given a particle
label, another particle in the same chaining cell is found using the
previous element as an index, the new element being zero if there are
no more charges in the chaining cell. For a more detailed explanation of this computer algorithm
we refer to the book by Hockney and Eastwood \cite{HandE}.

There is another problem associated with working on a lattice: the
interaction includes finite self-energies for the charges: because
each point charge is spread over several lattice sites, it interacts with
itself with finite energy. The self-energy in this context is a one particle potential energy that has
regular wells at positions relative to the lattice sites. This energy must be subtracted off, once again using (\ref{eqn:energy})
(with the sum taken over the sites over which each charge is spread) and (\ref{Greensfn}). This subtraction is
essential; without it, one finds that the charges become trapped within the lattice-induced potential wells at low temperatures, while
higher temperature results are also strongly affected by this potential.

\section{Local method} \label{localmethod}

\subsection{Continuum update} \label{continuum}

The novelty of the local algorithm is in the way the field configuration changes. When a charge moves, the field changes in the
vicinity of that charge so that Gauss' law is preserved. We first consider charge moves in the continuum limit to illustrate some
features of the dynamics. Because, in this subsection, we are not directly concerned with computations, we work with SI units and in three dimensions.

The electric field at time $t$ due to particle motion can be written as
\[ \mathbf{E}(\mathbf{r},t) = \mathbf{E}(\mathbf{r},t_{0}) -
\sum_{i}\frac{q}{\epsilon_{0}}\int d\mathbf{r}_{i}
\delta(\mathbf{r}-\mathbf{r}_{i}(t)) \]
where $\mathbf{E}(\mathbf{r},t_{0})$ is the purely longitudinal electric field at time
$t_{0}$ when the charges are in their starting positions. The electric field, according
to this equation, evidently varies locally as the charges move. Each charge leaves a trail along its path as pointed out by Levrel
and Maggs \cite{LevrelMaggs}.  If we take any closed surface around some charges at $t_{0}$, Gauss' law is
satisfied by the definition of $\mathbf{E}(\mathbf{r},t_{0})$. As time
increases, charges may enter or leave this closed surface, whereupon
the delta function ensures that flux is chopped out or added in at the
cutting point to preserve Gauss' law. If $F$ is the closed
surface 
\begin{multline*} -\frac{q}{\epsilon_{0}}\oint_{F}d\mathbf{S}\cdot\int d\mathbf{r}_{i}
\delta(\mathbf{r}-\mathbf{r}_{i}(t)) \\  = \left\{ \begin{array}{ll}
    \pm q/\epsilon_{0} & \mbox{if the charge cuts $F$ an odd number of times} \\ 0 & \mbox{otherwise} \end{array} \right. \end{multline*}
where the sign on the right-hand-side depends on whether the charge is originally inside (minus sign) or outside $F$.
The electric field trail term can be written as
\[ - \sum_{i} \frac{q}{4\pi\epsilon_{0}} \int d\mathbf{r}_{i}
    \nabla^{2}\left( \frac{1}{|\mathbf{r}-\mathbf{r}_{i}(t)|} \right)
    \]
in three dimensions. Using the identity \[\nabla^{2}\mathbf{V} =
\mathbf{\nabla}(\mathbf{\nabla}\cdot\mathbf{V}) -
\nabla\times\nabla\times\mathbf{V} \]
for vector field $\mathbf{V}$ we notice that, in accordance with the
Helmholtz theorem, the electric field can be split into longitudinal
and transverse parts
\[ \mathbf{E} = \mathbf{\nabla}\phi + \mathbf{\nabla}\times\mathbf{Q}
\]
where
\[ \phi(\mathbf{r}) = \sum_{i}
\frac{q}{4\pi\epsilon_{0}}\mathbf{\nabla}\cdot\int d\mathbf{r}_{i}
\left( \frac{1}{|\mathbf{r}-\mathbf{r}_{i}(t)|}\right) +
\phi(\left\{\mathbf{r}_{i}(t_{0})\right\}) \]
and 
\[ \mathbf{Q} = \sum_{i}\frac{q}{4\pi\epsilon_{0}} \mathbf{\nabla}\times\int d\mathbf{r}_{i}
\left( \frac{1}{|\mathbf{r}-\mathbf{r}_{i}(t)|}\right). \]
We can go further and write the scalar and vector potentials as 
\begin{multline*} \phi(\mathbf{r}) = \sum_{i}
\frac{q}{4\pi\epsilon_{0}}\int
\frac{d\mathbf{r}_{i}\cdot(\mathbf{r}-\mathbf{r}_{i}(t))}{|\mathbf{r}-\mathbf{r}_{i}(t)|^{3}}
\\ =  \sum_{i}
\frac{q}{4\pi\epsilon_{0}}\int d\mathbf{r}_{i}\cdot\mathbf{\nabla}_{i}
\left( \frac{1}{|\mathbf{r}-\mathbf{r}_{i}|}\right) +
\phi(\left\{\mathbf{r}_{i}(t_{0})\right\}) \end{multline*}
which is recognizable is the ordinary electric scalar potential while
\[ \mathbf{Q} = \sum_{i}\frac{q}{4\pi\epsilon_{0}}\int
\frac{d\mathbf{r}_{i}\times(\mathbf{r}-\mathbf{r}_{i}(t))}{|\mathbf{r}-\mathbf{r}_{i}(t)|^{3}}
\]
 which is the Biot-Savart law. So, the electric field at each time $t$
 is the Coulomb field due to the charges at time $t$ positions, and a
 circulation $\mathbf{\nabla}\times{\mathbf{Q}}$ where
 $\mathbf{\mathbf{Q}}$ is equivalent to the magnetic field produced by
 current-carrying wires along the trails left by the charges as they
 move - the current being $q/\mu_{0}\epsilon_{0}$.

We notice that if a charge makes a closed loop, there is no change in the longitudinal field as we should expect. Secondly, the transverse field is always zero for
the closed loop except on the loop itself, in accordance with $\mathbf{\nabla}\times{\mathbf{Q}}$ being proportional to the current
density in ordinary electromagnetism. When the path is not closed, this relationship breaks down; in terms of the current-carrying
wires analogy, the endpoints imply that charge is not conserved so there is an additional magnetic field contribution, the curl of which is identical to the Coulomb
  field of the charge at the endpoints. We have demonstrated, that pure transverse field updates occur entirely by the motion of
  charges in closed loops. 

\subsection{Off-lattice charge move update}

\begin{figure}
\begin{center}
\includegraphics[width=0.9\columnwidth,clip]{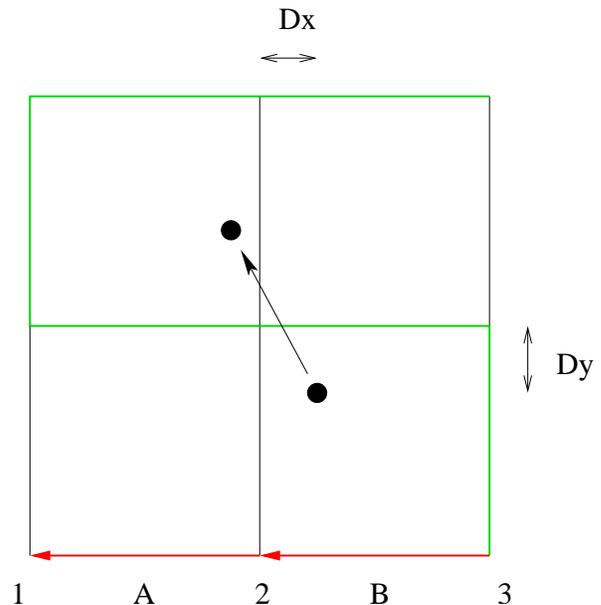}
\end{center}
\caption[Field update after a charge move and charge assignment to the vertices of the lattice.]{Charge update for charge move in neighborhood of a single
  plaquette. Completed link updates are shown in red. Those to be
  updated in green. The original charge position from the nearest vertex is given by $Dx$ and $Dy$ and the charge assignments on
  vertices $1$,$2$ and $3$ are initially given by $W^{y}_{-1}W^{x}_{-1}$, $W^{y}_{-1}W^{x}_{0}$ and $W^{y}_{-1}W^{x}_{1}$ respectively where the
  charge assignment weights $W$ are given in equation (\ref{eqn:weights}). If, when the charge moves, the change on vertex $i$
  is $\Delta q(i)$ then the link field denoted A is updated by $E_{A} \rightarrow E_{A}+(\Delta q(1)/a)$ and for link B $E_{B} \rightarrow E_{B}+((\Delta
  q(1)+\Delta q(2))/a)$ and so on. }
\label{fig:path}
\end{figure}

 The O(N) Monte Carlo charge update requires us to maintain the discretized version of Gauss' law, equation (\ref{latticeGauss}),
   at each charge move by changing a few links as possible for a given charge assignment function. The energy difference
   computation at each step from equation (\ref{eqn:energy2}) involves only the link fields that have changed. There are many possible
   ways of updating the electric field locally. When charges are completely confined to the lattice sites, then hopping of charge
   $q$ across a link requires us to alter the electric field $E$ on that link by $E\rightarrow E-q/a$ to preserve the flux
   constraint. The charge leaves a trail of electric field, as we have already seen in the continuum case. I have used the
   following field update procedure for off-lattice simulations \cite{RossettoMaggs}. Given the restrictions on charge motion and
   the assignment function, the number of vertices affected by the charge move must be either nine or twelve. Suppose that the
   nearest vertex to the charge $\mathbf{R}_{0}$ is the same before and after the move so that only nine vertices are affected;
   the other case is identical in principle. For each affected vertex $\mathbf{R}(p,q)\equiv\mathbf{R}_{0}+p\mathbf{x}+q\mathbf{y}$ with $p,q = -1,0,1$, the
change in the charge is $\Delta q(\mathbf{R}(p,q))$. Beginning at the vertex $\mathbf{R}(-1,-1)$, we trace out an $S$ shape,
updating the links along the path so that, at each site along the path, Gauss' law is satisfied. Figure~\ref{fig:path} shows the general idea. 


\subsection{Other features} \label{localfeatures} 

In continuum terms, the local update is responsible for generating an electric field of the form
\begin{equation} \mathbf{E} = -\mathbf{\nabla}\phi + \mathbf{\nabla}\times\mathbf{Q} + \mathbf{E}_{b} \label{eqn:field}\end{equation} 
with $\mathbf{Q}$ and uniform field $\mathbf{E}_{b}$ non-vanishing at finite temperature. By definition, the potential produces an electric
field with no net background field $\mathbf{E}_{b}$, but the local field update gives rise to fluctuations in this field. To see this, consider a charge that
hops from one lattice site to a neighboring site. The simplest local field update produces a field trail that would restore the
charge to its original position at zero temperature. If the uniform background were zero before the charge hopped between sites, then because the
background field is the sum over the vectors at each site divided by the total number of links there must such a field in the final state. This field can be spread over the lattice by further
field updates. From this kind of argument, we see that the background field that is generated depends on the distance of the
charges from their original positions. More precisely,  at time $t$, a single charge will be responsible for a uniform electric field that is proportional to $q(\mathbf{r}(0)
-\mathbf{r}(t))$ where $\mathbf{r}(0)$ is the original position of the charge $q$ and $\mathbf{r}(t)$ is its position at time
$t$. It should be emphasized that $\mathbf{r}$ is measured without any regard for the periodicity of the cell. In other words, the
electric field increases as a charge winds around the cell so the charge experiences a restoring force towards its original
position. With many charges, the uniform field is proportional to $\sum_{i}q_{i}(\mathbf{r_{i}}(0) -\mathbf{r_{i}}(t))$. So we
should expect that fluctuations in the uniform field will cause charges to lose sense of their original positions.

It has been argued that the local update can produce, even after averaging, an interaction of the more general type considered by Leeuw and Perram - a
periodic potential and a term that depends on the dipole moment of the cell \cite{RottlerMaggs} and which is equivalent to the
presence of a uniform electric field within the cell. We recall from Section~\ref{methods} that the periodic interaction is
recovered from a spherical cluster of identical cells by cladding the cluster with a perfect conductor. If the cladding has finite
dielectric constant $\epsilon$, a dipole term is introduced as in equation (\ref{LPdipole}). The dipole term in the cell is
bounded because the cell is finite. In contrast, the uniform field produced by the local is unbounded and does not depend on the
absolute dipole moment unless the dipole moment of the cell is zero at the beginning of the simulation. The argument
\cite{RottlerMaggs} is that at high temperatures, charge winding causes the uniform electric field to average out during the
simulation. At low temperatures though, the field does make a difference to the effective interaction after thermally averaging.

One can see that the uniform field makes no difference to thermal averages provided it is sampled independently of charge
positions. So one way of ensuring infinite $\epsilon$ is to introduce an independent background field update. This can be
implemented locally. It has the disadvantage of slowing the simulation down. A more efficient means is to keep track of the
uniform field locally and to subtract off its contribution to the energy. This can be done because the uniform field energy
decouples from other contributions even before thermal averaging. We shall investigate the influence of the uniform field further
in the next section.

\begin{figure}
\begin{center}
\includegraphics[width=0.9\columnwidth,clip]{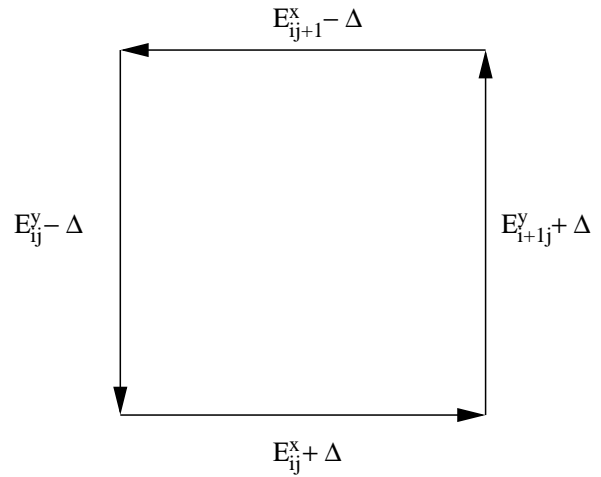}
\end{center}
\caption{Plaquette circulation update at lattice site (ij). The electric field divergence at each site is preserved.}
\label{fig:plaq}
\end{figure}

Although charge updates do sample transverse electric field configurations, we expect to have to sample these circulations
independently of the charge moves. For the circulation sampling, a plaquette is chosen at random from a uniform distribution over the
whole lattice. A number $\Delta$ is then selected from a uniform distribution over the range $[-\nu,\nu]$
where $\nu$ is chosen so that the move has a $50\%$ acceptance
rate. The circulation of the electric field around the chosen
plaquette is changed by altering each of the links of the plaquette by
$\Delta$ as shown in figure \ref{fig:plaq}. The energy difference
calculation therefore involves only four link variables.  


\subsection{Literature} \label{literature}
 
Monte Carlo with O(N) scaling by the introduction of transverse field degrees of freedom may have charges confined to the lattice
as introduced in \cite{RossettoMaggs,LevrelMaggs}, or allowed to move off-lattice with some smooth charge assignment onto the lattice
vertices \cite{RottlerMaggs} which we have followed in our simulations.

At first sight, we would expect that the transverse field
degrees of freedom would have to be allowed to relax across the whole
system for each particle move. Although the scaling of the speed
varies as order $N$ when only particle moves are considered, one would naively
expect order $N^{2}$ scaling when transverse field sampling is
included. It turns out, however, \cite{Maggs1} that one can sample the
transverse fields much less frequently than one would expect without
spoiling the results. This is a surprising result. In connection with this issue, Maggs and coworkers \cite{Review} have drawn
attention to the fact that the transverse electric field is sampled by charge moves without requiring independent plaquette
updates. In this paper we expand on this observation.

It has been found that for charges confined to a lattice, charges tend
to freeze in their original positions. Two papers are devoted to various ways of removing this obstacle in on-lattice
simulations \cite{LevrelMaggs,Duncan} by improving the efficiency of circulation updates or by spreading the charges over several sites. For off-lattice
simulations, this problem can be present but one can avoid it by taking the
step size to be small or by spreading charges over more sites. 

The local update method has been developed by Maggs and collaborators for molecular
dynamics \cite{RottlerMaggs2,Dunweg} and in work by Pasichnyk and D\"{u}nweg \cite{Dunweg}, the competitiveness of
local molecular dynamics was remarked on in a comparison with a global $O(N)$ method.

The method has been applied in various test cases in the original papers. More recently however, it has been used by Maggs to
carry out Monte Carlo simulations of a model of water \cite{Maggs2}. 

\section{Results of computer experiments} \label{simulations}


\subsection{The one-component plasma in two-dimensions.}

Our simulations focus on the one-component plasma in two dimensions. This is a system of particles, all with charge $q$,
on a fixed neutralizing background. In two dimensions the interaction
between the charges is logarithmic. The phase diagram of
this plasma depends only on the coupling $\Gamma$ where $2\pi\Gamma \equiv q^{2}\beta$. For $\Gamma=2$ the model can be solved exactly for
the free energy and the $n$-point correlation functions. At around $\Gamma=140$,
Monte Carlo simulations using periodic boundary conditions have
revealed the existence of a first-order transition from a liquid to a
crystalline phase. Most of the numerical work in the following focuses
on this model. Firstly, the exact solution allows us to perform an objective test of the accuracy of the lattice
based methods. Secondly, the plasma has long-ranged correlations for $\Gamma>2$ which form the basis for a stronger test of the
achievable accuracies. Thirdly, the model has a simple formulation and is suitable for making clean tests of other properties of
the new method including the effect of varying the rate of independent circulation updates.


   \subsection{Simulations of the 2D OCP} \label{2DOCP}

I have carried out Monte Carlo simulations of the one-component plasma
in two dimensions using the local electrostatic algorithm discussed at
length in the previous section and also a continuum method using the Lekner form for the pairwise interaction. The simulation cell
for the discrete methods is divided into $32\times 32$ plaquettes and each chaining cell is a  $2\times 2$ plaquette region.
The short range correction algorithm extends over a $6\times 6$ plaquette square with the
charge that is moved in the central chaining cell. I used the approximation to the Lekner potential (\ref{SRCmain}) as the
correcting potential. All the results in this subsection were taken with $120$ charges in the
simulation cell. The lattice spacing $a$ was chosen so that the number density of the particles was unity.  The rate of plaquette
updates was usually taken to be 40 for each charge move. Differences between the results obtained here and the correct thermal
averages arise from small differences in the potential energy for small charge separations because an approximation to the correct
potential is used (see appendix \ref{ewaldapprx}) and also from imperfect sampling when the relaxation times are large which
happens towards the lower temperature range we have investigated. If the approximation to the exact interaction were
insufficiently accurate, one could use the exact potential instead.

\begin{figure}
\begin{center}
\includegraphics[width=0.9\columnwidth,clip]{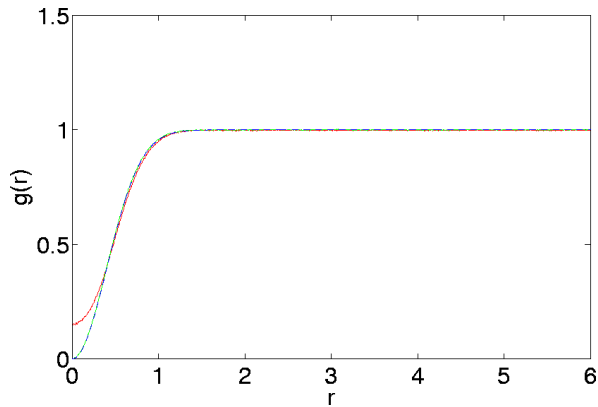}
\end{center}
\caption[Effect on pair correlation function of short range correction
  to interaction.]{Pair correlation functions for the 2D one-component plasma at
  $\Gamma=2$ with (red) and without (green) short-range potential
  correction. When the short-range potential is corrected the exact
  solution (also plotted) and simulation result overlap. }
\label{fig:gamma2}
\end{figure}

 For various values of the coupling $\Gamma$, I have computed the pair correlation function
$g(\mathbf{r_{1}},\mathbf{r_{2}})$ which is defined as
\begin{multline*} g(\mathbf{r}_{1},\mathbf{r}_{2}) \\ = (2\pi)^{2}\frac{1}{\pi^{N}}\frac{N!}{(N-2)!} \int \left(\prod_{i>2}^{N} d^{2}\mathbf{r}_{i}\right) P_{N}(\mathbf{r}_{1},\ldots,\mathbf{r}_{N}) \end{multline*}
where $P_{N}(\mathbf{r}_{1},\ldots,\mathbf{r}_{N})$ is the thermal probability distribution function for the 2D OCP with $N$ harmonically confined
charges with the number density set to one.  This quantity is proportional to the probability of finding charges in
the neighborhood of positions $\mathbf{r}_{1}$ and $\mathbf{r}_{2}$. Because the liquid, in the thermodynamic limit, is homogeneous and isotropic,
the pair correlation function depends only on $r\equiv |\mathbf{r}_{1}-\mathbf{r}_{2}|$. This definition for the pair correlation function has the property
that it tends to unity for large values of its argument. For $\Gamma=2$, Ginibre \cite{Ginibre} has found $g(r)$ exactly to be 
\begin{equation} g(r) = 1 - \exp(- \pi r^{2}). \label{GinibreEqn} \end{equation} This exact result exists because the partition
function for $N$ charges is identical to the normalization of a wavefunction of free fermions with Gaussian wavepackets. In
Figure~\ref{fig:gamma2}, the exact Gaussian curve is plotted with results from the Monte Carlo simulations.There is only weak
correlation for small charge separations; otherwise $g(r)$ is featureless. Both methods reproduce the correct behavior extremely
well; for the case of the lattice method, the correlation function is correct at small separations between the charges
provided the short range correction method is used. Without correcting the potential for small charge separations, the radial
density does not tend to zero as $r$ tends to zero, but is close to the exact solution for charge separations greater than about one particle spacing.

\begin{figure}
\begin{center}
\includegraphics[width=0.9\columnwidth,clip]{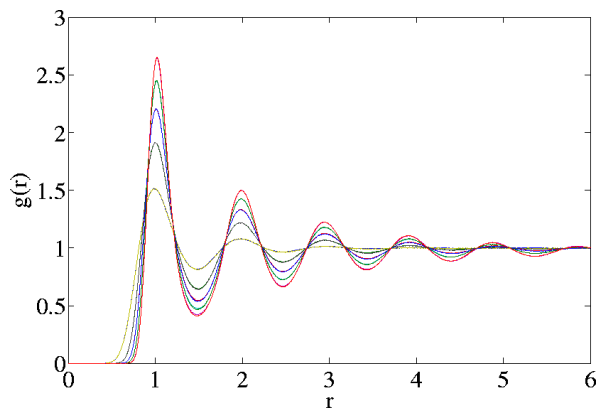}
\end{center}
\caption[Pair correlation functions for various different
  temperatures: a comparison of different simulation methods.]{Pair correlation function $g(r)$ for the 2D one-component
  plasma for $\Gamma$ equal to $20$,$40$,$60$,$80$,$100$. Results
  plotted for simulations with local update, and with the potential energy computed as a Lekner summation.}
\label{fig:rdfs}
\end{figure}

At higher values of $\Gamma$, we make a direct comparison of pair correlation functions from Monte Carlo data. There is excellent agreement
between the Monte Carlo methods for $\Gamma\leq 100$ as shown in Figure~\ref{fig:rdfs}. As we should expect, correlations become more
pronounced and longer-ranged as $\Gamma$ increases. For lower temperatures, discrepancies between the radial densities for
simulations using the Lekner potential and the local simulation appear. They are small at $\Gamma=100$
but become very pronounced for $\Gamma\sim 140$ which roughly corresponds to the reported onset of the crystalline
phase. Correlations are consistently weaker for the local method. 
 
\begin{figure}
\begin{center}
\includegraphics[width=0.9\columnwidth,clip]{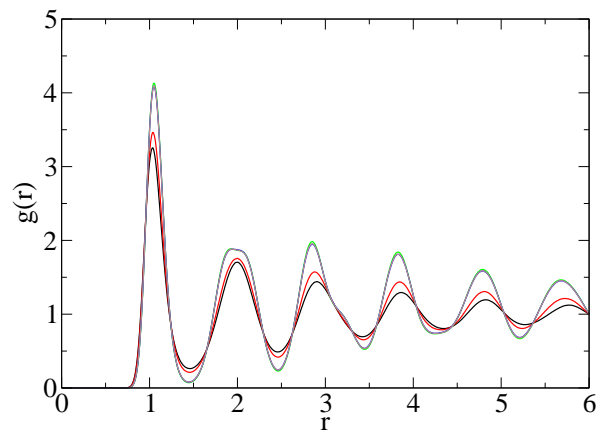}
\end{center}
\caption{Pair correlation function $g(r)$ for the 2D one-component
  plasma for $\Gamma=140$. Five curves are shown. The gray curve is for a global continuum method. The green and blue curves that
  match it were found using the local method with a short-range correction extending about $1/3$ of the total distance across the
  system from each charge. They are distinguished by either including independent uniform field sampling (green) or not
  (blue). The two less correlated curves (red and black) are respectively with and without uniform field sampling. They are less
  correlated than the other three curves because the corrections to the potential extend only half of the distance from each
  charge compared to the more correlated case.  }
\label{140rdf}
\end{figure}

By increasing the range of the short-range correction, the results for the local method can be brought arbitrarily close to those for the Lekner summation at the expense of
slowing down the energy computations. For example, if we extend the range of the short-range correction from $6\times 6$ to
$12\times 12$ plaquettes, we can obtain a perfect match between the Lekner correlation function and the local method result as shown in
Figure~\ref{140rdf}. 

Because the agreement between the Lekner potential and local method results is excellent even down to very low temperatures, we
rule out the possibility that the background field can make a significant contribution to the potential. As a further test, we
have included an explicit uniform field sampling term in our simulations. The results are shown in Figure~\ref{140rdf}. The pair
correlation functions match the Lekner result with or without the extra update when the $12\times 12$ correction region is used. A
difference does show up within the smaller correction region but the result with the extra update is less correlated than the
result without the field update. I believe that one cannot rule out the winding of the electric field at any temperature provided
the charges are mobile and that therefore, the uniform field will average out in most cases.


I have also computed the average energies of the charged particle system using the Lekner summation as a further test of the local
method. Of course, global energy calculations at each step defeat the object of the local algorithm, which is most efficient if we are only
interested in charge correlations. Anyway, the energies from this work agree temperature for temperature with results obtained
by Leeuw and Perram \cite{Leeuw2} to their degree of accuracy.


\subsection{Autocorrelation times of charges} \label{chargeautoc}

In this section and the following, we present various autocorrelation times denoted $\tau$ which are characteristic times of the autocorrelation
function
\[ A(k) = \frac{\langle O_{i}O_{i+k} \rangle - \langle O_{i}
  \rangle^{2} }{\langle O_{i}^{2} \rangle - \langle O_{i}
  \rangle^{2}} \]
where $O_{i}$ for $i=1,\ldots,N$ is a sequence of data obtained from the Monte Carlo simulation and the angled brackets denote a
  thermal average. These times provide a measure for
  the rate of relaxation of different observables over the course of the simulation. They depend on the update method, on the
  nature of the particles and their interaction and on the choice of $O$. We measure these using both a
  binning procedure \cite{Janke, Sokal} and by integrating the autocorrelation function. We find the numbers are consistent with
  one another, though the former method is preferable because it signals convergence to $\tau$. We present results in this section for two typical observables at various
  temperatures. After an equilibration process from a hot start, the
algorithm is allowed to adjust the maximum step size $\Delta \mathbf{r}$ that each charge can make from its original position so that the acceptance rate for charge
moves is $50\%$ \footnote{The choice of $50\%$ is arbitrary but conventional. It is possible that faster relaxation could be
  obtained with a different acceptance rate.}. The sampling is performed with this fixed maximum step size. A similar and
  simultaneous procedure is adopted for the plaquette update step in the local simulation.  Because we are concerned with charge
  observables in this section, we measure autocorrelation times in units of charge move trials; in other words independent
  transverse field updates are not included in the estimates.  

\begin{figure}
\begin{center}
\includegraphics[width=0.9\columnwidth,clip]{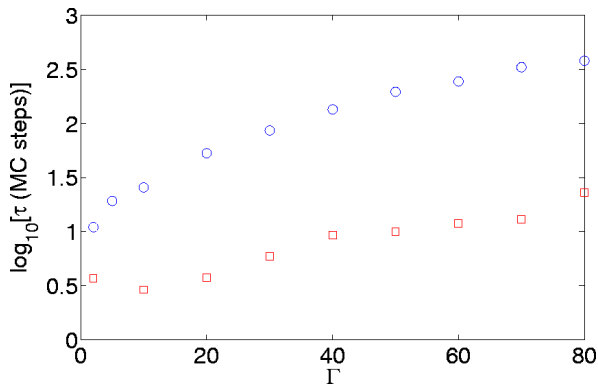}
\end{center}
\caption[Variation of autocorrelation times with temperature for two
  different simulation methods.]{Autocorrelation times for the energy on a logarithmic scale varying with $\Gamma$. The circles are data
  for the local method, squares for the Lekner simulation. }
\label{fig:Etau}
\end{figure}

Figure \ref{fig:Etau} shows a rise in the energy autocorrelation time as
$\Gamma$ increases; as the temperature is lowered, energy fluctuations
become more improbable so the step size  $\Delta \mathbf{r}$ falls to
retain the fixed acceptance rate. The autocorrelation time falls
coincidently. Over the range $\Gamma=2$ to $\Gamma=80$ there is
roughly a tenfold increase in $\tau$ for both the Lekner
and local methods. The autocorrelation times are consistently much
larger for the local method than for the Lekner method. That the local
simulation  $\Delta\mathbf{r}$ should be smaller than that for the
Lekner simulation is to be expected on the following grounds. In the
absence of charges, a single charge has to overcome an energy barrier,
on average, to make a move using the local method. This is because it
has to move in its own field which relaxes locally to maintain Gauss'
law. This can be seen most easily in the case of a single charge
confined to a lattice vertex. If we ignore field circulation changes
which only produce fluctuations about the average that we consider
here, a step in the position of the charge takes a single link field
from $E$ to $E- (q/a)$ where $E=q/4a$ (in 2D) on average; the energy
difference is then $q^{2}/4>0$. In contrast, a single charge in a
periodic cell in the Lekner simulation has no barrier to overcome to
make a move on average. The qualitative observation that the energy barrier for a given
step size is higher in the local method than in the global method is
true regardless of the number of charges present in the cell.


\begin{figure}
\begin{center}
\includegraphics[width=0.9\columnwidth,clip]{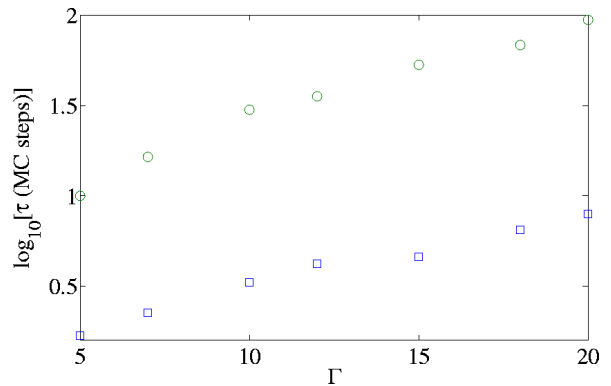}
\end{center}
\caption[Comparison of structure factor relaxation time for global and
  local simulation methods.]{Variation of $\log_{10}(\tau)$ of the structure factor for
  $\mathbf{k}$ vector $(10,6)$ (close to the first reciprocal lattice
  vector) with inverse temperature $\Gamma$ in the liquid phase which
  exhibits short-range correlations. The local simulation (circles)
  relaxes more slowly than the global simulation method with a Lekner
  potential.}
\label{fig:TCFtau}
\end{figure}

I have also computed the structure factor for $k$ vector close to the
first reciprocal lattice vector for our two different simulation methods. As correlation develops in the
sample, a peak forms in the structure factor at reciprocal lattice
vectors. The structure factor at these values of $k$ are the slowest
to relax so we have concentrated on high temperatures compared to the
reported phase transition temperature to ensure good results. Once again, we find that $\tau$ is
much larger for the local simulation method than for the global method as
can be seen from Figure~\ref{fig:TCFtau}, which uses a logarithmic scale to
represent the autocorrelation times.  


\subsection{The relative speed of the local algorithms} \label{speed}

 In previous works, the $O(N)$ scaling of the local method has been pointed out. But, such
scaling cannot be of much use if the scaling prefactor or, in other words, the absolute times are prohibitively large even for
small systems. In this section, we shall see that in comparison with the Lekner summation simulations, the local method fares very
well even when the differing autocorrelation times are included. The autocorrelation time provides a measure for the efficiency
with which simulation methods can sample configuration space. Specifically, if a single run gives a sequence of $n$ values of some
observable with variance $\sigma_{O_{i}}^{2}$, the variance of the data about the ensemble average for that observable is given
by \cite{Janke,Sokal}
\begin{equation} \sigma_{\mbox{\={O}}}^{2} = \sigma_{O_{i}}^{2}\frac{2\tau}{n}. \label{variance} \end{equation}  
We use the autocorrelation time of some observable to find the number of sweeps required to bring the variance about the ensemble
mean down to some value and hence obtain the absolute simulation times taking into account the differing charge relaxation
times. Different observables have different autocorrelation times in general but to make a comparison of absolute speeds in this way we have to choose one. We have chosen
$S(\mathbf{k}=\mathbf{G})$, because, as a charge correlation function, it is a quantity that is both interesting and which the
local algorithm is particularly suited to computing. Also, the reciprocal lattice vector is likely to give the slowest relaxation
of charge observables, so the test will provide something of an upper bound on times. It should be emphasized that the 
autocorrelation time for sufficiently high temperatures or sufficiently large system sizes, is independent of system size
provided it is measured in Monte Carlo sweeps. So, the comparison made in this section does not affect the scaling of the algorithms. This is
not the case close to phase transitions, which would require a separate analysis, or for very small system sizes.  

I have measured the time taken for both the local and Lekner algorithms to complete a given number of particle moves for two
system sizes $N=120$ and $N=1000$. This time depends, of course, on the code, the compiler and the machine that is used so we
don't expect the numbers presented here to demonstrate any more than trends in the absolute and relative speeds and rough
estimates for the absolute speed. The Lekner method scales as $O(N^{2})$ but it has the advantage that it is easy to implement and
the time taken for a charge move should be smaller than in $O(N)$ molecular dynamics methods for typical system sizes.

The code with local updates has a short-range potential correction routine with a $6\times 6$ plaquette region within which the
correction is performed. The simulation cell is a $32\times 32$ mesh. The number of plaquette updates per particle move is
maintained at $40$ so that the mobility is close to maximum as we shall see in Section \ref{transautoc}. The cell size is chosen
so that the charge density is one.

 The raw speed of the local code before including the correlation in the data is obviously much greater than that of the global
code because the overheads are so much smaller when just a few links need to be changed. We also find that the times scale as $N$
for the Lekner code as expected. The time interval for a fixed number of steps in the local code does vary with $N$ only because
the number of particles captured by the short-range correction routine increases with $N$. In fact, the short-range correction makes a very
significant contribution to the speed: for a cell with $200$ charges, switching off the correction routine halves the time
interval. When the number of plaquettes increases holding the number of particles per plaquette constant, the speed of the local
update code does not change significantly. 

In the previous section, the variation of $\tau$ for $S(\mathbf{k}=\mathbf{G})$ with temperature was given. We have also
measured the mean and variance of the structure factor for each temperature from the data. Suppose we want to estimate
$S(\mathbf{k}=\mathbf{G})$ to within $1\%$. We obtain the required number of configurations $n$ from equation
(\ref{variance}). 

\begin{table}
\begin{tabular}{|l|l|l|l|l|l|l|l|}                 \hline\hline

$\Gamma$               &   5.0   &  7.0   &  10.0  &  12.0  &  15.0  &  18.0  &  20.0  \\  \hline 

$\tau$     &  1.7    & 2.25   &  3.3   & 4.2    &  4.6   &  6.5   &  7.9   \\
/MC sweeps          &  10.0   & 16.4   &  30.0  & 35.6   &  53.2  &  68.4  &  94.4  \\  \hline  

No.MC sweeps for   &  7.7    & 11.0   &  17.8  & 23.7   &  27.8  &  42.8  &  53.9  \\ 
$1\%$ s.d./ $10^{3}$ & 45.6    & 80.7   &  162.6 & 199.4  &  325.6 & 446.0  & 641.9  \\  \hline

Simulation time  & 10.7     & 15.4   &  24.9  & 33.2   &  38.9  &  60.0  & 75.0   \\    
120 charges/min   & 4.9      & 8.6    &  17.3  & 21.2   &  34.7  &  47.5  & 68.3   \\  \hline

Simulation time  & 12.6     & 18.1   &  29.2  & 39.0   &  45.7  &  70.5  & 88.7   \\   
1000 charges/hr   & 3.2      & 5.7    &  11.5  & 14.0   &  23.0  & 31.5   & 45.3   \\  \hline

\end{tabular}
\caption[Comparison of speeds of Lekner and local simulation methods
  for different temperatures.]{Table of speeds of the Lekner and local
  simulation methods taking into account the fact that their autocorrelation
  times differ. We used a 1004MHz, AMD Athlon x86 64 bit processor with Linux operating system and suitably optimized gcc
  compilation. The upper figure in each box is for the Lekner summation Monte Carlo, the lower number for the local method.}
\label{table:speed}
\end{table}

Table \ref{table:speed} gives, for seven temperatures, the autocorrelation time of the first
reciprocal lattice peak of the structure factor (which has already appeared in Figure~\ref{fig:TCFtau}), the variance
in the configuration values about the mean and hence the number of sweeps required to
get the error of $S(\mathbf{k}=\mathbf{G})$ to $1\%$ of the mean value. From our measurements of the
speed of each piece of code, we estimate the minimum amount of time the global and local simulations should
be allowed to run to obtain data of this quality for $N=120$ and $1000$. The effect of including the autocorrelation times is
contained in the two lowest rows. The local code remains the faster of the two, for both system sizes and for all
temperatures but inclusion of the relaxation times has narrowed the gap considerably. For $\Gamma=20$ and
$N=120$ the necessary times are brief (as we would expect at these fairly high temperatures compared to the phase transition) and
similar. The scaling of the speeds with the number of charges implies a larger difference for $N=1000$ but the absolute times are
becoming inconveniently large.


   \subsection{Transverse field relaxation} \label{transautoc}

The unique feature of the local simulation method is the introduction of transverse field degrees of freedom. These degrees of
freedom should scale roughly with $N$, the number of charges in the system to preserve the accuracy of results. One might expect that
the transverse electric field should be allowed to relax all the way across the system at each charge move, with the result that
the algorithm should scale as $O(N^{2})$. In practice, we find that a very low rate of independent plaquette updates $O(1)$ is all that is
required per charge move to obtain good sampling at intermediate temperatures $2<\Gamma<140$ and, at higher temperatures, plaquette updates
can be switched off altogether. At lower temperatures, we find that one must sample the transverse field at a higher rate per
charge move. 

In this section, we investigate these properties. To do this, we have measured the autocorrelation time $\tau$ of the
background uniform electric field as the rate of plaquette updates is changed. The uniform field does not depend directly on the
plaquette circulation; it varies as charges move by the local update method as discussed in Section~\ref{methods}, so this quantity
provides a measure for the charge relaxation times that is particularly straightforward to investigate compared to relaxation
times of charge correlation functions. 

\begin{figure}
\begin{center}
\includegraphics[width=0.9\columnwidth,clip]{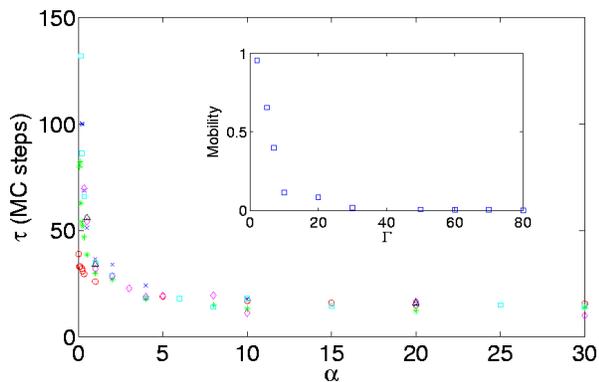}
\end{center}
\caption[Effect of plaquette update rate and temperature on
  $\mathbf{q}=0$ field autocorrelation time.]{Variation of the integrated autocorrelation time of the uniform background
  field with plaquette update rate measured in average number of
  plaquette moves per particle move. Results for four different
  temperatures are shown: $\Gamma=2$,$5$, $20$,$40$, $60$ and $140$
  (circles,stars,crosses,squares,diamonds and triangles respectively). The definition of $\alpha$ is given in the main text. The
  inset shows the gradient of RMS distance of particles from their original positions varying with $\Gamma$.}
\label{fig:BKGtau}
\end{figure}

Figure~\ref{fig:BKGtau} shows how $\tau$ varies when the plaquette update rate is varied. The horizontal axis in the figure is 
\[ \frac{\mbox{Rate of plaquette updates}}{\mbox{Rate of particle updates}} \equiv \alpha. \]  
We find that, as $\alpha$ increases from zero for intermediate temperatures, $\tau$ decreases sharply and reaches a
vale. Over the temperature range $5\leq\Gamma\leq 140$ that we have investigated, the behavior of $\tau$ is so weakly
dependent on temperature that the data almost collapse. A similar behavior is seen in the mobility of charges,(defined here as the
gradient taken from the plot of RMS distances of charges from their original positions against simulation time measured in Monte Carlo sweeps). The mobility is close to maximum
for a very low rate of plaquette updates: just one update per particle move on average although there are $32\times 32$ plaquettes
and roughly one particle for nine plaquettes. We have also looked directly at the autocorrelation times of different modes of the
transverse field with the same temperature independence even close to the transition. At higher temperatures $\Gamma<20$, we see a significant change in the behavior of
$\tau$ for the uniform field. The same figure~\ref{fig:BKGtau} shows data for $\Gamma=2$ for which $\tau$ falls below the lower temperature data
at $\Gamma=5$ at small $\alpha$. We have also been able to obtain the $\alpha=0$ point for $\Gamma=2$ which is too large at lower
temperatures to obtain accurately. This high temperature behavior matches the qualitative result that good averages are obtained
at high temperatures $\Gamma\leq 2$ without plaquette updates.   
 
We can understand this behavior qualitatively as follows. We have already pointed out that in the absence of plaquette updates,
the electric field is altered by the motion of charges, which leave trails of altered field along their paths. The effect of the
trail on each charge is to restore it to its original position. So, at least, at low temperatures, we expect charges to be trapped
in their original positions with a correspondingly long autocorrelation time of the uniform field which oscillates with the
charges. When the electric field is allowed to relax by changing the circulation around plaquettes, the string is smeared across the system
and the restoring force on each charge falls on the average and hence the charges become mobile. There must be some limiting
mobility at a fixed temperature. We expect that smearing of the trails can also take place purely by the motion of charges - an
effect which is presumably enhanced by spreading charges over more lattice sites. But this smearing of trails by charges is
contingent on at least some independent plaquette updates. At high temperatures, charges can overcome the potential barriers
around their initial positions so $\tau$ for the uniform field is small.

\begin{figure}
\begin{center}
\includegraphics[width=0.9\columnwidth,clip]{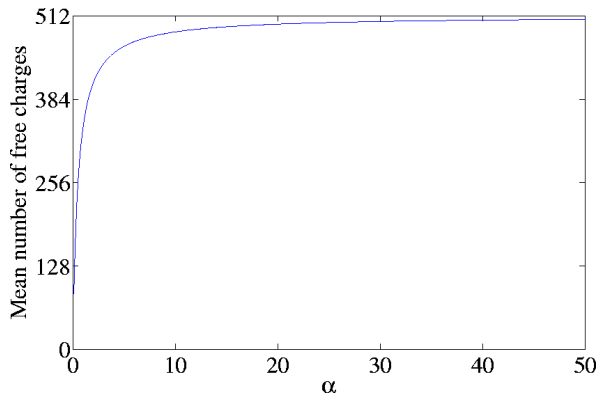}
\end{center}
\caption{$\langle N_{F}\rangle$ as a function of $\alpha$ showing the
  rapid approach to $512$ where there are $1024$ plaquettes in total.}
\label{fig:modelmean}
\end{figure}

We would like to understand why the maximum mobility is attained with such a low update rate at intermediate temperatures. We
shall see that a simple probability model for the update process exhibits a similar feature. Suppose
there are $N$ plaquettes, each one inhabited, on average, by a charge. At each step, the probability
of choosing a plaquette is $P_{p}$ and the probability of choosing a charge $P_{c}$. The ratio
$P_{p}/P_{c}\equiv\alpha$. So
\[ \begin{array}{lr} P_{p} = \dfrac{\alpha}{\alpha+1}  &  P_{c} = \dfrac{1}{\alpha+1}. \end{array} \]
A plaquette is said to be free if a charge on that plaquette can move freely (it does not leave an electric
field trail). Otherwise, it is trapped. The number of free plaquettes is $N_{F}$. The following
processes are possible
\begin{enumerate}
\item Plaquette move frees plaquette: $N_{F}(t_{i-1})+1 = N_{F}(t_{i})$
\item Plaquette move traps plaquette: $N_{F}(t_{i-1})-1 = N_{F}(t_{i})$
\item Particle move on trapped plaquette: $N_{F}(t_{i-1}) = N_{F}(t_{i})$
\item Particle move on free plaquette: $N_{F}(t_{i-1})-1 = N_{F}(t_{i})$.
\end{enumerate}
The fourth rule that a charge moving on a free plaquette causes the
plaquette to become trapped has its basis in the fact that a moving
charge leaves a line of electric field when it moves.
The probability distribution function of the number of free plaquettes satisfies a master equation
\[ \frac{\partial P(N_{F},t)}{\partial t} = J(N_{F}) - J(N_{F}+1,t) \]
where
\[ J(N_{F}) = \frac{\alpha}{\alpha+1}\frac{N-N_{F}+1}{N}P(N_{F}-1,t) - \frac{N_{F}}{N}P(N_{F},t).
 \]
We solve for the stationary (time-independent) probability distribution in the usual way \cite{Gardiner}. Form the summation
\[ \sum_{i=0}^{M-1} \left[ J(i+1) - J(i) \right] = J(M) - J(0). \]
The boundary conditions on the rates give $J(0)=0$ and hence
\[ P(N_{F}) = \frac{\alpha}{\alpha+1}\frac{N-N_{F}+1}{N_{F}} P(N_{F}-1). \]
We compute the mean number of free plaquettes $\langle N_{F}\rangle$ as the ratio $\alpha$ is
changed. The mean number of free plaquettes is essentially equivalent to the number of free charges
and is therefore indicative of the mobility. We find that $\langle N_{F}\rangle$ tends rapidly to a maximum of half of $N$ as $\alpha$ is
increased (see Figure~\ref{fig:modelmean}). Around $10$ plaquette updates per particle move are sufficient to bring
$\langle N_{F}\rangle$ within $5\%$ of maximum just as is the case for mobilities in the simulations
presented above. Of course, this model is hugely oversimplified. There are many ways in which the simulation updates differ from the processes in
this model. For example, the simulation uses a Monte Carlo test to carry out updates, charges are
spread over four plaquettes and we have simulated at a particle density of about $1$ particle to $9$
plaquettes (neglecting spreading). Also, a charge can erase a trail by moving along it: a trapped plaquette can be freed up by
charge motion. Nevertheless, the model captures the important feature of the changing mobility: if more than one plaquette is
freed per particle move, $N_{F}$ increases until limited by the unfreeing operation. We therefore expect the limiting number of free plaquettes to be
half of the total.    


We turn now to the behavior at low temperatures around the transition temperature. We find that a high rate of plaquette updates
is necessary to achieve the same degree of correlation one finds with the Lekner summation method. For $\Gamma=120$ and $140$, I
have obtained pair correlation functions at different values of $\alpha$ in the range $\alpha=2$ to $\alpha=100$. All the
$\Gamma=120$ results are identical and match the pair correlation function using the Lekner Monte Carlo method. The $\Gamma=140$
results differ, with convergence onto the Lekner result for around $\alpha=40$. I have not investigated this further but it is
worth pointing out that the slower convergence to the correct result as $\alpha$ is varied happens close to the transition
temperature where one would expect charge relaxation times to be longer than in the fluid phase. There is no remarkable behavior
close to the transition temperature in the uniform field relaxation time nor in the transverse field relaxation time. For
practical purposes at such low temperatures, the unbiased plaquettes updates are inefficient and it may be sensible to use a worm
update to sample transverse degrees of freedom \cite{Review} which involves the creation of a pair of equal and opposite charges,
at intervals during the simulation, which wander around the simulation cell and eventually annihilate.

We conclude this section with three observations. The first concerns the relationship between the mobility (defined above) and the autocorrelation
time of the uniform field $\tau$. We have found that as $\tau$ falls, the mobility rises. For low $\alpha$, this is consistent
with the picture of charges trapped by their electric field trails. However, the mobility is also temperature dependent as can be
seen in the inset plot of Figure~\ref{fig:BKGtau}. At low temperatures, the mobility is small although the uniform field may relax
quickly. So, $\tau$ appears to be independent of the mobility. Notice however, that the greatest temperature change in the
mobility and in $\tau$ occurs for $\Gamma<20$ whereas for $\Gamma>20$ both are only weakly temperature dependent, so they are more
(anti)correlated for low $\alpha$.

The second observation is that the curve of uniform field $\tau(\alpha)$ depends on the number of charges $N$ and the number
of plaquettes $P$ only in the ratio $P/N$. If the number of plaquettes increases, holding the number of charges constant, the
autocorrelation time increases because the fraction of plaquettes updated per particle move drops. When $N$ increases in
proportion to $P$, the data collapses onto the curve in Figure~\ref{fig:BKGtau}.

Finally, the results described in this section lead to the conclusion that if charges can overcome the impetus to retrace the electric field trails
they produce (which will typically be at high temperatures), then thermal averages will be electrostatic averages even in the
absence of independent plaquette updates. After all, the Metropolis algorithm guarantees that averages from computer generated configurations
will converge to thermal averages (which in this case are electrostatic averages) provided that all of configuration space can be sampled in principle. In practice, reasonable
mobilities can be achieved at high temperature with no plaquette updates. At lower temperatures, a few plaquette updates are
required to overcome bottlenecks while electric field trails are altered by charge motion and charges can cooperate in freeing
other charges provided they are dense enough that charge clouds overlap to some extent. 

Although charge dynamics depends on electric field circulation and vice versa, the factorization of transverse and longitudinal energies ensures
that these contributions to the fields are importance sampled independently and simultaneously. Relaxation of transverse fields across the periodic
sample at each charge step is unnecessary because we are interested in averages over many configurations. It is sufficient that relaxation can occur
over a number of charge steps. At the end of the simulation with no information discarded, one can 'project' onto each link transverse field to find it has been sampled
adequately. Each transverse field is a harmonic oscillator for which good thermal averages are easily obtained and even with a
separate transverse degree of freedom of each link, most of the work goes into sampling charge configurations.

If this conclusion is correct, one would expect that the charge dynamics introduced in Section~\ref{continuum} in the presence of
thermal noise should be sufficient to recover electrostatic averages. Unfortunately, the evolution of the electric field
configuration is not a Markov process so further analysis is difficult.


\section{Summary and conclusions} \label{conclusions}

We have made a study of a local Monte Carlo method developed by Rottler and Maggs \cite{RottlerMaggs} for simulating systems of
charged particles. The unrefined method is limited in the accuracy it can achieve, because the electric field is discretized onto a
lattice. One must remove self-energy contributions due to charge spreading and correct the short-range interaction. All these can
be done locally, so the scaling of the method is $O(N)$ where $N$ is the number of charges.

1. Simulations of the 2D one-component plasma have been carried out in one of the first applications of the local off-lattice
  Monte Carlo method. With a short-range interaction correction extending a distance of about a tenth of the length $L$ of each
  simulation cell from each charge, excellent results are obtained for the pair correlation function for $\Gamma\leq 100$. By
  making the method less local results can be obtained to lower temperatures. For example, with short-range correction a distance
  of $L/5$ from each charge, good results can be obtained at least down to the transition temperature $\Gamma\sim 140$. 

2. Although there is a non-vanishing uniform electric field during our local simulation which is a consequence of the charge move
   update, we find that this produces no significant deviation from results obtained from an Ewald potential for which the dipole
   term is zero. It is not certain from this study whether this result will generalize to all systems containing many mobile charges but I suspect
   that the electric field will indeed average out because charge winding, which gives an electric field that cannot be inferred from the
   particle positions, is always important. 

3. In order to obtain accurate results at low temperatures ($\Gamma\sim 140$), a much larger rate of plaquette updates is
  necessary than at lower temperatures. This is correlated with the fluid-crystal transition. Plaquette updates are completed much
  more quickly than charge moves so this effect does not substantially alter the speed though of course charge autocorrelation times are larger at lower temperatures so simulations should be longer anyway. One might
  introduce a worm update to sample the transverse electric field more efficiently at low temperatures though it would seem to be
  unnecessary otherwise, at least for the 2D OCP.

4. The charge autocorrelation times for the local method are much larger than for global methods. This is because there is, on
  average, a larger energy barrier to charge motion when Gauss' law is maintained locally. 

5. The autocorrelation times should be included in calculations of the effective speed. When this is done for system sizes that
  are typically simulated, we find that the gap between the local method and the $O(N^{2})$ global method is considerably narrowed
  compared to the raw speeds, but that the local method remains significantly faster and the gap widens as the number of charges
  increases. We have not studied molecular dynamics methods which in some cases have $O(N)$ scaling. Preliminary results by Pasichnyk and
  D\"{u}nweg \cite{Dunweg} suggest that molecular dynamics adapted from the local Monte Carlo we have discussed here is a promising alternative to
  the P$^{3}$M method. But, we expect Monte Carlo to have the advantage of relative ease of implementation and smaller scaling
  prefactor. The most significant contributions to the prefactor are charge assignment, local field updates and especially
  corrections to the lattice potential. Except at low temperatures, plaquette updates are not a significant factor in the speed.

6. We also looked at the question of how the introduction of extra degrees of freedom can result in a faster method when naively we
   would expect the transverse degrees of freedom should be allowed to relax across the simulation cell for each charge move. Charge mobility for intermediate
   temperatures is seen to increase sharply as the rate of plaquette updates is increased from zero. We have provided a simple
   model that, from a reasonable premise, reproduces this qualitative behavior. For high temperatures $\Gamma\leq 2$, the charge
   mobility is large even when independent plaquette moves are switched off. The fact that plaquette updates can be switched off
   at high temperatures strongly suggests that importance sampling of the simple transverse degrees of freedom can be accomplished
   entirely by charge moves, provided they are sufficiently mobile.   

7. We have introduced a continuum version of the charge move electric field update which allows us to see more rigorously that charge
   moves alter the transverse electric field. Following from the observation in the previous point, an interesting problem is to
   carry out the average over electric fields using this dynamics alone and hence recover electrostatic averages.

8. In an appendix, we have showed that a thermal average over Maxwell electrodynamics for a system of charges gives the same charge
  correlation functions one would obtain from electrostatics alone. As is well-known, the magnetic field has no effect on
  correlation functions. Also, the transverse part of the electric field averages out as in the original discussion by Maggs and Rossetto.

\acknowledgments
I would like to thank both A. C. Maggs and M. A. Moore for useful
discussions at various stages of this work. I would also like to acknowledge financial support from the EPSRC. 


\appendix

\section{Approximation to the Ewald potential in two dimensions} \label{ewaldapprx}

In this section, we find a good approximation to the Ewald potential
in two dimensions for small charge separations.
Consider 2 charges $q$ in a rectangular periodic cell with side
lengths $L_{x}$ and $L_{y}$. The whole system is charge neutral
because a neutralizing charge is smeared uniformly over the
cell. Leeuw and Perram \cite{Leeuw} have found that the total
energy can be expressed as
\begin{multline} 2\pi U = \frac{q^{2}}{4}\sum_{\mathbf{n}}\sum_{i,j=1}^{2}\star
E_{1}\left(\alpha^{2}(\mathbf{r}_{ij}+\mathbf{n})^{2}\right) \\ + \frac{q^{2}}{2\pi
A}\sum_{\mathbf{k}\neq0}
    \frac{\exp(-\pi^{2}\mathbf{k}^{2}/\alpha^{2})}{\mathbf{k}^{2}}\left[ 1 - \cos(2\pi\mathbf{k}\cdot\mathbf{r}_{12}) \right] \\ - \frac{q^{2}}{2}\left(\gamma + \ln \alpha^{2}A\right) - \frac{q^{2}\pi}{\alpha^{2}A} 
\label{eqn:ewald}\end{multline}
where the $\star$ means that we should not include the charge self
energy in the summation over $\mathbf{n} = (n_{x},n_{y})$ and $n_{i}$
are integers. The area of the system is $A = L_{x}L_{y}$ and $\gamma$
is Euler's constant. The energy does
not depend on constant $\alpha$ but the rate of convergence of the
summations does.
This is the energy of the interaction between the charges and their
periodic images, interaction of the charges with the background charge
and the self-energy of the background. We would like to subtract off
the background contributions. The interaction between two charges
without images is just $(q^{2}/2\pi)\ln(1/r)$. So we compute
\[  U_{\mbox{\small bkg}} = \lim_{r\rightarrow0}{U + \dfrac{q^{2}}{2\pi}\ln r} \] 
because the interaction of the charges with their images are
guaranteed to vanish in the limit. Hence the interaction between the
charges is given by $U_{int} = U - U_{bkg}$. We find
\begin{multline} 2\pi U_{int} = \frac{q^{2}}{2}
  E_{1}(\alpha^{2}\mathbf{r}^{2}) \\ + \frac{q^{2}}{2\pi
A}\sum_{\mathbf{k}\neq0}
    \frac{\exp(-\pi^{2}\mathbf{k}^{2}/\alpha^{2})}{\mathbf{k}^{2}}\left[ -1 + \cos(2\pi\mathbf{k}\cdot\mathbf{r}) \right] \\ - \frac{q^{2}}{2}(\gamma + \ln \alpha^{2}) 
\end{multline} where the distance between the charges is denoted $r$.   
We approximate the exponential integral for small $r$
\[ E_{1}(\alpha^{2}r^{2}) = -\gamma - 2\ln \alpha r + \alpha^{2}r^{2}
+ \dots \]
We now turn to the $k$ space summation which, for small $r$ is
\[ -\frac{q^{2}}{4\pi
A}\sum_{\mathbf{k}\neq0}
    \frac{\exp(-\pi^{2}\mathbf{k}^{2}/\alpha^{2})}{\mathbf{k}^{2}}\left(
    2\pi\mathbf{k}\cdot\mathbf{r} \right)^{2} -
    \frac{q^{2}}{2}(\gamma + \ln \alpha^{2}). 
\]
The cross-terms in $\mathbf{k}\cdot\mathbf{r}$ vanish because the
summation is over positive and negative integers. We can also remove a
factor of $r^{2}$ because the terms in the summation are
identical. Then we use the identity
\[ \sum_{\mathbf{k}}\exp{-\pi^{2}\mathbf{k}^{2}/\alpha^{2}} =
\frac{\alpha^{2}A}{\pi}\sum_{\mathbf{n}}\exp{-\alpha^{2}A\mathbf{n}^{2}} \]
and approximate the exponential for large $A$, retaining the terms to
order $A$.
We find that the short range correction energy $U_{SRC}$, which approximates $U_{int}$, is
\begin{equation} U_{SRC} = -\dfrac{q^{2}}{2\pi}\ln r + \dfrac{ q^{2}r^{2}}{4A} \label{SRC} \end{equation}
the terms depending on $\alpha$ cancel. The calculation in three dimensions is given in an appendix to \cite{Fraser}.

\begin{figure}
\begin{center}
\includegraphics[width=0.9\columnwidth,clip]{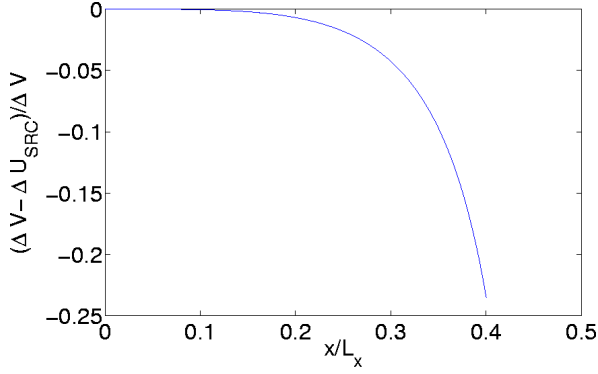}
\end{center}
\caption{Plot of the normalized error between energy differences calculated using the Lekner potential and the short range
  interaction potential (\ref{SRC}). We have used parameters appropriate to the simulations we have performed, with $32\times 32$
  square plaquettes covering the cell and lattice spacing $a=0.342$. The energy differences are for charge steps in the $x$
  direction through a distance of one-half of the lattice spacing. $L_{x}$ is the width of the periodic cell in the $x$ direction.}
\label{fig:error}
\end{figure}

We compare this result with the Ewald potential. The figure (\ref{fig:error}) shows how the approximation to the short range
correction given by equation (\ref{SRC}) differs from the Lekner interaction $V$ from equation (\ref{Lekner}) as the charge separation increases. The measure we use is
\[ \frac{\Delta V(x) - \Delta U_{SRC}(x)}{\Delta V(x)} \]
where $\Delta U_{SRC}(x)$ and $\Delta V(x)$ are energy differences as a charge steps outwards by half a lattice spacing from separation $x$. The error is
within $1\%$ for charge separations of $1/5$ of the cell width $L_{x}$ and for the highest accuracy low temperature simulations we
performed, up to $2\%$.


\section{Electrostatics from thermal electrodynamics} \label{maxwell}

In this part we wish to demonstrate that the argument Maggs and
Rottler have made for their rather unphysical energy function holds
also for full Maxwell electrodynamics. That is to say that even though
the charges do not influence one another by instantaneous action at a
distance, they nevertheless generate thermodynamic averages as though
they do. This result has already been found by Alastuey and Appel \cite{Alastuey} but I believe it would be useful to present the
argument in this context. The main difficulty is in finding an energy functional that
includes the constraints that depend on the static configuration of
charged particles. 

The charges are not subject to an external field but are influenced by
other charges in our hypothetical box in a heat bath. Since the fields
are dynamical quantities we must include them in our discussion. So,
the Lagrangian for a system of $N$ relativistic charges $q_{i}$ for
$i=1,2,\ldots,N$ interacting electrodynamically is
\[ L = \sum_{i=1}^{N} -m_{0i}\sqrt{1-\mathbf{v}_{i}^{2}} + \int
d^{3}\mathbf{r}\left(-\frac{1}{4}F_{\mu\nu}F^{\mu\nu} + J_{\mu}A^{\mu} \right) \]
where $\{\mathbf{v}_{i}\}$ are the particle velocities and $\{m_{0i}\}$ their rest masses.
$F_{\mu\nu}\equiv\partial_{\mu}A_{\nu} - \partial_{\nu}A_{\mu}$ is the
electromagnetic field tensor defined in terms of $4$-vector potential
$A_{\mu}$. The $4$-vector current has components $J^{0}$ which is the
charge density and spatial components in  $3$-vector $\mathbf{J} =
\sum_{i}\delta(\mathbf{r} - \mathbf{r}_{i})q_{i}\mathbf{v}_{i}$. We have taken $c=1$. 

The form of the Lagrangian ensures that the homogeneous Maxwell
equations are obeyed \footnote{The electric field is $E_{i}\equiv -F_{0i}$ and the magnetic field is $B_{i} =
  (1/2)\epsilon_{ijk}F_{jk}$}
\[ \partial_{\mu}F_{\nu\sigma} + \partial_{\sigma}F_{\mu\nu} +
\partial_{\nu}F_{\sigma\mu} = 0 \left\{ \begin{array}{r} \mathbf{\nabla}\cdot\mathbf{B} = 0 \\
  \mathbf{\nabla}\times\mathbf{E} =
  -\frac{\partial\mathbf{B}}{\partial t}\end{array}\right. \]
The first step in getting a Hamiltonian is to find the conjugate
  momenta from the Lagrangian. For the charges, the momenta are
\[ \mathbf{p}_{i} = \frac{\partial L}{\partial \mathbf{v}_{i}} =
  \frac{m_{0i}\mathbf{v}_{i}}{\sqrt{1-\mathbf{v}_{i}^{2}}} +
  q_{i}\mathbf{A}_{i} \] where $\mathbf{A}_{i} =
  \mathbf{A}(\mathbf{r}_{i})$ and for the fields  
\[ \Pi^{\mu} = \frac{\partial \mathcal{L}}{\partial_{0}
  \mathbf{A}_{\mu}} = -F_{0\mu} \]
where $\mathcal{L}$ is the Lagrangian density. We notice that $\Pi^{0}$ vanishes because $F^{\mu\nu}$ is antisymmetric. The equations of motion are
\[ \partial_{\mu}F^{\mu\nu} + J^{\nu} = 0.  \]

If we go straight ahead and find the Hamiltonian for the fields and
the interaction term
\[ H_{F} = \int d^{3}\mathbf{r} \,  \mathbf{\Pi}\cdot\dot{\mathbf{A}} - \int
d^{3}\mathbf{r} \left(-\frac{1}{4}F_{\mu\nu}F_{\mu\nu} + J_{\mu} A^{\mu}\right) \]
we get 
\[ H_{F} = \int d^{3}\mathbf{r} \, \left( \frac{1}{2}\left(\mathbf{B}^{2} + \mathbf{E}^{2}\right) - \mathbf{J}\cdot\mathbf{A}\right) \]
in terms of the electric and magnetic fields having used an equation of motion after partial integration to write 
\[ \mathbf{\Pi}\cdot\dot{\mathbf{A}} = F_{0i}F_{0i} + A_{0}J^{0}. \]

In this form, we cannot perform an average over the electric field because it is constrained by Gauss' law 
\[ \mathbf{\nabla}\cdot\mathbf{E} = J^{0} \]
which is formally an equation of motion although it is a static constraint on the field. We shall now render the Hamiltonian in a
form that brings out the important constraints. To do this, we first notice that the electric field can be split into rotational and irrotational
parts, exactly as we did in the introduction. Using the same
notation as before
\[ \mathbf{E} = \mathbf{E}_{\|} + \mathbf{E}_{\bot}. \]
We also recall that the Lagrangian for electrodynamics is gauge invariant. We can
fix the gauge to reduce the arbitrariness in $\mathbf{A}$. For our
purposes it is useful to choose the so-called Coulomb gauge
$\mathbf{\nabla}\cdot\mathbf{A}=0$. This can always be done. The
reason for choosing this gauge is that the splitting of the electric
field into transverse and longitudinal parts is mirrored in the vector
potential in the following nice way
\[ \begin{array}{cc} \mathbf{E}_{\|} = -\mathbf{\nabla}A^{0} &
  \mathbf{E}_{\bot} = -\dfrac{\partial\mathbf{A}}{\partial t}. \end{array}\]
We see that Gauss' law in terms of
$A_{0}$ is Poisson's equation which has the solution
\[ A^{0}(\mathbf{r},t) = \int d^{3}\mathbf{r}' \  
\frac{J^{0}(\mathbf{r}',t)}{4\pi|\mathbf{r} - \mathbf{r}'|} \]
which is the instantaneous Coulomb interaction. The transverse field
is independent of the Gauss' law constraint. It is also significant
that, as before, the electric field energy splits into independent
transverse and longitudinal energies
\[ \frac{1}{2}\int d^{3}\mathbf{r} \   \mathbf{E}^{2} = \frac{1}{2}\int d^{3}\mathbf{r} \  
\mathbf{E}_{\bot}^{2} + \frac{1}{2}\int d^{3}\mathbf{r}d^{3}\mathbf{r}' \  
\frac{J^{0}(\mathbf{r})J^{0}(\mathbf{r}')}{4\pi|\mathbf{r} -
  \mathbf{r}'|}. \]
Henceforth we denote the longitudinal field energy by $V$.
Now that we have put the longitudinal dependence into $A^{0}$ we would
like to have a momentum variable that is independent of $A^{0}$. We
choose $\mathbf{\Pi}_{\bot} = \mathbf{\Pi} - \mathbf{\nabla}A^{0}$
which must satisfy the constraint
$\mathbf{\nabla}\cdot\mathbf{\Pi}_{\bot}=0$ that follows directly from
the gauge constraint. We must get the Hamiltonian from this
new momentum. For the field  quantities
\[ H_{F} = \left(\int d^{3}\mathbf{r} \  \mathbf{\Pi}_{\bot}\cdot\dot{\mathbf{A}}\right) - L_{F}, \]
where $L_{F}$ is the Lagrangian excluding the field-independent parts, so that the complete energy function (including the free particle and
interaction terms) is 
\begin{multline*} H = \sum_{i} \sqrt{ \left(\mathbf{p}_{i} - q_{i}\mathbf{A}_{i}\right)^{2}+m_{0i}^{2}}  \\ + \int d^{3}\mathbf{r} \left[
  \frac{1}{2}\mathbf{\Pi}_{\bot}^{2} +
  \frac{1}{2}(\mathbf{\nabla}\times\mathbf{A})^{2} + V \right].
\end{multline*}
The partition function in the canonical ensemble with  inverse
temperature $\beta$ is
\begin{multline*} Z(\beta) = \int \prod_{i}d^{3}\mathbf{p}_{i} \int
\prod_{i}d^{3}\mathbf{r}_{i} \int \mathcal{D}\mathbf{\Pi}_{\bot}
\delta(\mathbf{\nabla}\cdot\mathbf{\Pi}_{\bot}) \\ \times \int \mathcal{D}\mathbf{A}
\delta(\mathbf{\nabla}\cdot\mathbf{A}) \exp(-\beta H) \end{multline*}
and the limits of the particle momentum integrations are infinite.
The field momentum integration factorizes from the rest. Then we make the change of variables
$\mathbf{P}_{i} = \mathbf{p}_{i} - q_{i}\mathbf{A}_{i}$ for each
particle. The Jacobian for this transformation is unity because the
vector potential does not depend on the momenta. So, the integral over
the vector potential also factorizes and we are left with
\[ Z(\beta) =
Z(\beta)_{\mathbf{P}}Z(\beta)_{\mathbf{\Pi}}Z(\beta)_{\mathbf{A}} \int
\prod_{i}d^{3}\mathbf{r}_{i} \exp(-\beta V) \]
which, up to a configuration independent factor is the electrostatic
partition function. This is essentially a generalized version of the
well-known Bohr-van Leeuwen theorem (see, for example \cite{vanvleck}).

\end{document}